\documentclass[prb,twocolumn,amsmath,citeautoscript,showpacs]{revtex4}
\usepackage{graphicx}
\usepackage{bm}
\newcommand* {\vek}[1]{{\ensuremath{\bm{\mathrm{#1}}}}}
\newcommand* {\pp}{\vek{p}}
\newcommand* {\ee}{\ensuremath{\mathrm{e}}}
\newcommand* {\eps}{\ensuremath{\varepsilon}}
\newcommand* {\frack}[2]{{\Ts\frac{#1}{#2}}}

\newcommand* {\Ts}{\textstyle}

\newcommand* {\SSs}{\scriptscriptstyle}

\newcommand* {\expectk}[1]{\ensuremath{\langle {#1} \rangle}}
\usepackage{array}
\newcolumntype {s}[1]{@{\hspace{#1}}} 

\begin{document}

\title{Oscillatory multiband dynamics of free particles: The ubiquity of
\emph{zitterbewegung\/} effects}

\author{R. Winkler}
\affiliation{Department of Physics, Northern Illinois University,
DeKalb, IL 60115}
\affiliation{Kavli Institute for Theoretical Physics, University of
California, Santa Barbara, CA 93106}

\author{U. Z\"ulicke}
\affiliation{Institute of Fundamental Sciences and MacDiarmid Institute
for Advanced Materials and Nanotechnology, Massey University,
Private Bag 11~222, Palmerston North, New Zealand}
\affiliation{Kavli Institute for Theoretical Physics, University of
California, Santa Barbara, CA 93106}

\author{J. Bolte}
\affiliation{Institut f\"ur Theoretische Physik, Universit\"at Ulm,
Albert-Einstein-Allee~11, D-89069 Ulm, Germany}

\date{\today}

\begin{abstract}
  In the Dirac theory for the motion of free relativistic electrons,
  highly oscillatory components appear in the time evolution of
  physical observables such as position, velocity, and spin angular
  momentum. This effect is known as \emph{zitterbewegung}. We
  present a theoretical analysis of rather different Hamiltonians
  with gapped and/or spin-split energy spectrum (including the
  Rashba, Luttinger, and Kane Hamiltonians) that exhibit analogs of
  \emph{zitterbewegung\/} as a common feature. We find that the
  amplitude of oscillations of the Heisenberg velocity operator
  $\vek{v}(t)$ generally equals the uncertainty for a simultaneous
  measurement of two linearly independent components of $\vek{v}$.
  It is also shown that many features of \emph{zitterbewegung\/} are
  shared by the simple and well-known Landau Hamiltonian describing
  the dynamics of two-dimensional (2D) electron systems in the
  presence of a magnetic field perpendicular to the plane. Finally,
  we also discuss the oscillatory dynamics of 2D electrons arising
  from the interplay of Rashba spin splitting and a perpendicular
  magnetic field.
\end{abstract}

\pacs{73.21.-b, 71.70.Ej, 03.65.Pm}

\maketitle

\section{Introduction and overview}
\label{sec:intro}

The Dirac equation~\cite{dir28, sak67, tha92} was derived to obtain
a relativistic generalization of Schr\"odinger's approach to quantum
physics that describes the dynamics of single-electron quantum
states. While it served as an important stepping stone towards a
more complete description of quantum-electrodynamic effects, Dirac
theory has occasionally been regarded with some suspicion. In
particular, the effect of \emph{zitterbewegung\/}~\cite{sch30}
showed that solutions of the Dirac equation exhibit peculiarities
that are inconsistent with classical intuition in a more fundamental
way than nonrelativistic quantum physics. The
\emph{zitterbewegung\/} is an oscillatory dynamics of observables
induced by the Dirac equation, with a frequency of the order of
$2mc^2/\hbar$, where $m$ is the electron mass, $c$ is the speed of
light and $\hbar$ is the Planck constant. The amplitude of
oscillations in a particle's position is of the order of the Compton
wave length. Subsequently, \emph{zitterbewegung\/} attracted some
interest as a possible way to understand the intrinsic magnetic
moment of the electron~\cite{hua52,bar81}. Later, on the level of
fundamental physics, the advent of quantum field theory obviated the
need to discuss relativistic quantum theory in terms of a
first-quantized, Schr\"odinger-type theory. Present interest in the
Dirac equation ranges from hadronic physics~\cite{ben03} over
lattice gauge theory~\cite{smi02} to recent efforts~\cite{sch02} to
incorporate relativistic effects into quantum-chemistry
calculations.

A Dirac-like dynamics causing analogs of \emph{zitterbewegung\/} was
also predicted for electrons moving in crystalline
solids~\cite{fer90, zaw06}, in particular for narrow-gap
semiconductors~\cite{zaw05a}, carbon nano\-tubes~\cite{zaw05b},
graphene sheets~\cite{kat06}, tunnel-coupled electron-hole
bilayers~\cite{shm95} and superconductors~\cite{lur70}. All
these systems are characterized by having the relevant electron
excitations grouped into two bands separated by a nonzero energy gap
so that their energy spectrum is similar to the spectrum of the
Dirac Hamiltonian. A recent study~\cite{sch05a} of two-dimensional
(2D) electron systems in inversion-asymmetric semiconductor
heterostructures showed the presence of an oscillatory motion
analogous to \emph{zitterbewegung\/} arising from spin splitting of
the energy levels. The spin splitting corresponds to an energy gap
that vanishes for momentum $p \rightarrow 0$. A similar situation
occurs for electronic excitations in the bulk of an ideal graphene
sheet~\cite{kat06}.

These findings indicate the need to understand
\emph{zitterbewegung\/}-like effects on a more general level. In
Ref.~~\onlinecite{cse06}, the authors presented a general formula
for the Heisenberg position operator $\vek{r} (t)$ in systems that
can be described by effective $2\times 2$ Hamiltonians~\cite{2by2}.
In the present work, we have investigated the
oscillatory dynamics of Heisenberg observables such as position
$\vek{r}(t)$, velocity $\vek{v}(t)= d\vek{r}/dt$, orbital angular
momentum $\vek{L}(t)$, and spin $\vek{S}(t)$ in a variety of
qualitatively different models that describe the motion of free
(quasi-)particles. Besides the Dirac Hamiltonian, we have studied
three Hamiltonians frequently used in semiconductor physics to
describe the dynamics of (quasi-free) Bloch electrons in the
vicinity of the fundamental gap, the Rashba~\cite{byc84},
Luttinger~\cite{lut56}, and Kane~\cite{kan57} Hamiltonians. A number
of striking features emerge quite generally in all these models,
thus illustrating remarkable similarities between time evolutions
generated by rather different Hamiltonians. We suggest that these
common features can be used to extend the concept of
\emph{zitterbewegung\/} to a broader class of quantum Hamiltonians
for free (quasi-)particles. Our analysis shows that this generalized
notion of \emph{zitterbewegung\/} is manifested, in addition to the
oscillatory unitary time evolution of observables, also by
uncertainty relations characterizing the measurement of such
observables. These two aspects turn out to be closely related. In
particular, they can be described, for each of the models considered
here, by the same set of parameters. Also, we identify the typical
scales (lengths, velocities, and frequencies) that characterize
\emph{zitterbewegung\/}-like oscillatory motion. We emphasize that
this extended notion of \emph{zitterbewegung\/} is entirely based on
quantum mechanical concepts. In an alternative, semiclassical
approach one would identify a \emph{zitterbewegung\/} relative to a
suitable classical dynamics as, e.g., in Ref.~~\onlinecite{bol04}.
In some cases the conclusions will be different from those obtained
within the present approach. The most general aspects of our study
can be summarized as follows:

(i) An oscillatory motion occurs in the time evolution of free
(quasi-) particles when the energy spectrum of the corresponding
Hamiltonian $H$ is characterized by one or several energy gaps.
Besides the Dirac model, an important example are Bloch electrons in
solids~\cite{fer90,zaw06}, whose quantum dynamics are described by
effective free-particle Hamiltonians that incorporate the effect of
the periodic lattice potential.

(ii) In the case of two-band models (e.g., the Dirac, Rashba, and
Luttinger models), \emph{zitterbewegung\/}-like effects are
generally characterized by an amplitude operator $\vek{F}$ and a
frequency operator $\hat{\omega} (\vek{p})$. These two quantities
enter the expression for the velocity operator in the Heisenberg
picture, which can be decomposed as $\vek{v} (t) = \bar{\vek{v}} (t)
+ \tilde{\vek{v}} (t)$, where the mean part is
\begin{subequations}
  \label{eq:gen:vel}
  \begin{equation}
    \label{eq:gen:vel:av}
    \bar{\vek{v}} (t) =  \frac{\partial H}{\partial\vek{p}} - \vek{F}
  \end{equation}
  and the oscillating part is
  \begin{equation}
    \label{eq:gen:vel:osc}
    \tilde{\vek{v}} (t) = \vek{F} \, \ee^{-i \hat{\omega} (\vek{p}) \, t}
    = \ee^{i \hat{\omega} (\vek{p}) \, t} \, \vek{F} \quad .
  \end{equation}
\end{subequations}
Here, $\hbar\hat{\omega} (\vek{p})$ is related to the energy
difference between states having the same momentum $\vek{p}$, but
belonging to different subspaces (i.e., energy bands) of the
Hamiltonian. The operator $\vek{F}$, which anticommutes with
$\hat{\omega} (\vek{p})$, determines the magnitude of oscillations
in the velocity components but also enters the expression for the
mean part. We can integrate Eq.\ (\ref{eq:gen:vel}) to get the
Heisenberg position operator that can be decomposed in the same way,
$\vek{r} (t) = \bar{\vek{r}} (t) + \tilde{\vek{r}} (t)$, where
\begin{subequations}
  \label{eq:gen:posH}
  \begin{eqnarray}
  \label{eq:gen:pos:av}
  \bar{\vek{r}} (t) & = & \vek{r} + \bar{\vek{v}} t
  + \vek{F} \:\frac{1}{i \hat{\omega} (\vek{p})} \quad, \\
  \tilde{\vek{r}} (t) & = &
  - \vek{F} \: \frac{\ee^{-i \hat{\omega} (\vek{p}) t}}
  {i \hat{\omega} (\vek{p})} \quad .
\end{eqnarray}
\end{subequations}
Similarly, we get the time derivative of $\vek{v} (t)$,
\begin{equation}
  \label{eq:gen:accH}
  \dot{\vek{v}} (t) =  i \hat{\omega} (\vek{p}) \, \vek{F} \,\,
  \ee^{- i \hat{\omega} (\vek{p}) t} \quad .
\end{equation}
The operators $\vek{F}$ and $\hat{\omega} (\vek{p})$ govern also the
oscillations in the Heisenberg time evolution of the orbital angular
momentum operator $\vek{L} (t)$, and the spin operators $\vek{S}
(t)$. In systems with more than two bands (Kane and Landau-Rashba
models), more than one characteristic frequency and amplitude
operator can appear.

(iii) For each model describing an oscillatory multiband dynamics of
free particles, the components of the velocity operator $\vek{v}(t)$
do not commute. This can be written as an uncertainty relation that
takes the form (apart from a prefactor of order one)
\begin{equation}
  \label{eq:gen:vel:unc}
  \Delta v_j \, \Delta v_k \ge \tilde{v}^2
  \qquad (j \ne k),
\end{equation}
where $\tilde{v}$ is the amplitude of the oscillatory motion, see
Eq.\ (\ref{eq:gen:vel:osc}). The uncertainty relations
(\ref{eq:gen:vel:unc}) are an integral part of our
analysis~\cite{von97}.

(iv) The velocity operator $\vek{v}(t)$ does not commute with the
Hamiltonian. Although we discuss the motion of free (quasi-)particles,
the components of $\vek{v}(t)$ are not constants of the motion, see
Eq.\ (\ref{eq:gen:accH}). On the other hand, momentum $\vek{p}$
is always a constant of the motion. This implies that none of the models
discussed here provides a simple relation between momentum $\vek{p}$
and velocity $\vek{v}$.

(v) The counterintuitive properties of $\vek{r}(t)$ and $\vek{v}(t)$
arise because $\vek{r}(t)$ mixes different subspaces $\mathcal{H}_j$
that are associated with the different bands in the energy spectrum of
$H$. Thus we can interpret \emph{zitterbewegung\/}-like phenomena as
an interference effect. In the case of two-band models, one can replace
$\vek{r}$ by the part $\bar{\vek{r}}$ that leaves the subspaces
$\mathcal{H}_\pm$ associated with the `$+$' and `$-$' bands
separately invariant,
\begin{equation}
  \label{eq:dirac:pos:proj}
  \bar{\vek{r}} = P_+ \, \vek{r} \, P_+
  + P_- \, \vek{r} \, P_- \quad ,
\end{equation}
where $P_\pm$ are projection operators onto these subspaces. The
result coincides with the mean part $\bar{\vek{r}} (t)$ of
$\vek{r}(t)$ introduced in Eq.\ (\ref{eq:gen:pos:av}), i.e.,
\emph{zitterbewegung\/}-like effects are removed by the projection
(\ref{eq:dirac:pos:proj}). This result can be understood from a
different perspective by analyzing the amplitude operator $\vek{F}$.
We get
\begin{equation}
  \label{eq:dirac:F:proj}
  \vek{F} P_+ = P_- \vek{F}
  \quad\mbox{and}\quad
  \vek{F} P_- = P_+ \vek{F} \, ,
\end{equation}
i.e., $\vek{F}$ maps states associated with the `$+$' band onto
states associated with the `$-$' band and vice versa. An alternative
definition of $\bar{\vek{r}} (t)$ is obtained by applying the
inverse unitary transformation to $\vek{r}(t)$ that makes $H$
diagonal. The same techniques can also be applied to $\vek{v} (t)$
to obtain $\bar{\vek{v}} (t)$ given in Eq.\ (\ref{eq:gen:vel:av}).
The components of $\bar{\vek{v}} (t)$ commute; hence they can be
measured simultaneously [unlike Eq.~(\ref{eq:gen:vel:unc})]. They
also commute with the Hamiltonian so that they are constants of the
motion.

(vi) In every case considered, \emph{zitterbewegung\/}-like
phenomena are manifested also by oscillations of the orbital angular
momentum $\vek{L}(t)$ and spin $\vek{S}(t)$. At the same time, the
\emph{total} angular momentum $\vek{J}$ does \emph{not\/} oscillate
as a function of time. As expected for a model of a free particle,
$\vek{J}$ is a constant of the motion, i.e., it commutes with the
Hamiltonian. From a different perspective, this implies that the
oscillations of $\vek{L}(t) = \vek{r}(t) \times \vek{p}$ and
$\vek{S} (t)$ must cancel each other, which is possible only if the
oscillations of $\vek{r}(t)$ and $\vek{S}(t)$ have a common origin.
For the Rashba Hamiltonian, the oscillatory motion of $S_z(t)$
corresponds to the well-known and experimentally
observed~\cite{cro05} spin precession in the effective magnetic
field of the Rashba term.

The following Sections~\ref{sec:dirac}--\ref{sec:kane} are devoted
to a detailed discussion of \emph{zitterbewegung\/} effects arising
in systems whose time evolution is governed by the
Dirac~\cite{tha92}, Rashba~\cite{byc84}, Luttinger~\cite{lut56}, and
Kane~\cite{kan57} Hamiltonians. Remarkable formal similarities
between the oscillatory behavior of observables in these models are
established, as outlined above. Next we show in Sec.~\ref{sec:lan}
that the familiar Landau model of 2D electrons subject to a
perpendicular magnetic field~\cite{lan30} exhibits essentially all
the features attributed to the extended notion of
\emph{zitterbewegung\/} in previous sections. We finish our case
studies in Sec.~\ref{sec:lanrash} by investigating the quantum
oscillatory dynamics of 2D electrons arising from the interplay of
Rashba spin splitting and a perpendicular magnetic
field~\cite{ras60}. Conclusions and a summary of open questions are
presented in Sec.~\ref{sec:concl}. For easy reference, we provide a
number of relevant basic formulae in the Appendix.

\section{Dirac Hamiltonian}
\label{sec:dirac}

Our discussion of \emph{zitterbewegung\/} for the Dirac Hamiltonian
follows, for the most part, Ref.~~\onlinecite{tha92}. We include
this section with an overview of the effect's salient features to
provide a reference frame and notation for our following discussion
of solid-state analogies.

The Dirac Hamiltonian $H_D$ describes a free relativistic electron
or positron. It can be written in the form
\begin{equation}
  \label{eq:dirac}
  H_D = c \, \vek{\alpha} \cdot \pp + \beta \, m c^2 \quad ,
\end{equation}
where
\begin{equation}
\vek{\alpha} = \left(\begin{array}{ls{0.6em}l}
    0 & \vek{\sigma} \\ \vek{\sigma} & 0
  \end{array}\right) \, , \qquad
\beta = \left(\begin{array}{cs{0.6em}c}
    \openone_{2\times 2} & 0 \\ 0 & - \openone_{2\times 2}
  \end{array}\right) \, ,
\end{equation}
and $\vek{\sigma}$ is the vector of Pauli spin matrices. (Here we
assume magnetic field $B = 0$. See Ref.~~\onlinecite{bar85} for the
generaliziation to finite $B$.) The energy eigenvalues of $H_D$ are
$E_\pm (\vek{p}) = \pm \Lambda_D$, where
\begin{equation}
  \label{eq:dirac:en}
  \Lambda_D = \sqrt{m^2 c^4 + c^2 p^2} .
\end{equation}
Note that $H_D^2 = \Lambda_D^2$. In the Schr\"odinger picture, the
velocity operator reads
\begin{equation}
  \label{eq:dirac:velS}
  \vek{v} = \frac{i}{\hbar} [H_D, \vek{r}]
  = \frac{\partial H_D}{\partial\vek{p}}
  = c \,\vek{\alpha} \quad ,
\end{equation}
so that the components of $\vek{v}$ have the two discrete
eigenvalues $\pm c$. In the Heisenberg picture, we get
Eq.~(\ref{eq:gen:vel}) with
\begin{equation}
  \label{eq:dirac:F}
  \vek{F} = \frac{\partial H_D}{\partial \vek{p}}
  - \frac{c^2\,\vek{p}}{H_D} \;,
  \hspace{2em}
  \hat{\omega} (\vek{p}) = \frac{2 H_D}{\hbar} \quad .
\end{equation}
The operator $\vek{F}$ mediates a coupling between states with
positive and negative energies; see below. The oscillatory part
$\tilde{\vek{v}} (t)$ of $\vek{v}(t)$, given in Eq.\
(\ref{eq:gen:vel:osc}), describes the \emph{zitterbewegung\/}. The
frequency of the \emph{zitterbewegung\/} is (at least) of the order
of $\omega_D \equiv 2mc^2 / \hbar$. Integrating $\vek{v} (t)$ yields
the position operator $\vek{r} (t)$ in the Heisenberg picture, see
Eq.\ (\ref{eq:gen:posH}), which contains again the quickly
oscillating term $\ee^{-i \hat{\omega} (\vek{p}) t}$. The
oscillatory time dependence is similar to the motion of a
nonrelativistic particle in the presence of a magnetic field [see
Eq.\ (\ref{eq:lan:acc}) and discussion in Sec.~\ref{sec:lan}]. An
illuminating discussion of \emph{zitterbewegung\/} based on a
numerical calculation of the time evolution of wave packets can be
found in Ref.~~\onlinecite{tha04}.

It turns out \cite{tha92} that orbital angular momentum $\vek{L} =
\vek{r} \times \vek{p}$ and spin $\vek{S}$, which is defined as
  \begin{equation}
    \label{eq:dirac:spin}
    \vek{S} = -\frac{i\hbar}{4} \: \vek{\alpha} \times \vek{\alpha}
    = \frac{\hbar}{2}\left(\begin{array}{cc}
        \vek{\sigma} & 0 \\ 0 & \vek{\sigma}
      \end{array}\right) \, ,
  \end{equation}
show the phenomenon of \emph{zitterbewegung}, too. For the orbital
angular momentum, we have
\begin{subequations}
  \label{eq:gen:angH}
  \begin{equation}
    \label{eq:gen:orbangH}
    \vek{L}(t) = \vek{r} (t) \times \vek{p}
    = \vek{L} + \vek{F} \times \vek{p} \:
    \frac{1 - \ee^{-i \hat \omega(\vek{p}) t}}{i \hat{\omega} (\vek{p})} \, .
  \end{equation}
  The time evolution of spin in the Heisenberg picture reads
  \begin{equation}
    \label{eq:gen:spinH}
    \vek{S} (t) = \vek{S} - \vek{F} \times \vek{p} \:
    \frac{1 - \ee^{-i \hat{\omega} (\vek{p}) t}}{i \hat{\omega} (\vek{p})} \, .
  \end{equation}
  Thus it follows from Eqs.\ (\ref{eq:gen:orbangH}) and
  (\ref{eq:gen:spinH}) that the total angular momentum $\vek{J} =
  \vek{L} + \vek{S}$ does not oscillate as a function of time,
  \begin{equation}
    \vek{J} (t) = \vek{J} = \vek{L} + \vek{S} \quad ,
  \end{equation}
\end{subequations}
which reflects the fact that $[\vek{J}, H_D] =0$.

We can estimate the magnitude of \emph{zitterbewegung\/} by
evaluating the square of $\tilde{\vek{v}}(t)$
(Ref.~~\onlinecite{sch30}). This yields
\begin{equation}
  \label{eq:dirac:vel:osc}
  \tilde{v}^2 (t) = \frac{c^2 \, (2\Lambda_D^2 + m^2 c^4)}{\Lambda_D^2} \, ,
\end{equation}
i.e., $\tilde{v}^2$ varies between $3c^2$ in the nonrelativistic
limit and $2c^2$ in the relativistic limit. (Note that, although
$\tilde{v}^2 > c^2$, no measurable velocity exceeds $c$.) On the
other hand, the components of the velocity operator $\vek{v}$ do not
commute. Equations (\ref{eq:dirac:velS}) and (\ref{eq:dirac:spin})
imply that
\begin{subequations}
  \label{eq:dirac:vel:comunc}
  \begin{equation}
    \label{eq:dirac:vel:com}
    [v_j, v_k] = \frac{4ic^2}{\hbar} \: \eps_{jkl} \, S_l \quad .
  \end{equation}
  Diagonalizing this equation yields the uncertainty relation for $j
  \ne k$
  \begin{equation}
    \label{eq:dirac:vel:unc}
    \Delta v_j \, \Delta v_k \ge c^2
    = \big(\frack{1}{2} \, \omega_D \lambdabar_D \big)^2 \, ,
  \end{equation}
\end{subequations}
where $ \lambdabar_D = \hbar / (mc)$ is the Compton wave length.
Thus both the magnitude and the uncertainty of the
\emph{zitterbewegung\/} are given by $c^2$
(Ref.~~\onlinecite{von97}). We can also estimate the spatial
amplitude of the \emph{zitterbewegung\/} using the decomposition
$\vek{r}(t) = \bar{\vek{r}} (t) + \tilde{\vek{r}} (t)$, see Eq.\
(\ref{eq:gen:posH}). We get for the oscillating part
\begin{equation}
  \label{eq:dirac:pos:osc}
  \tilde{r}^2 (t) = \frac{\hbar^2 c^2 \, (2\Lambda_D^2 + m^2 c^4)}
  {4 \Lambda_D^4} ,
\end{equation}
i.e., in the nonrelativistic limit, the amplitude of
\emph{zitterbewegung\/} is approximately $\lambdabar_D$, and it is
given by the de Broglie wave length $\lambdabar_B = \hbar/p$ in the
relativistic limit.

It is well-known \cite{tha92} that \emph{zitterbewegung} is caused
by a coupling between the states with positive energies
(``particles'', subspace $\mathcal{H}_+$) and negative energies
(``antiparticles'', subspace $\mathcal{H}_-$). Thus one can
eliminate the oscillations of $\vek{r} (t)$ by projecting $\vek{r}$
on $\mathcal{H}_\pm$ as in Eq.\ (\ref{eq:dirac:pos:proj}) and the
result coincides with Eq.\ (\ref{eq:gen:pos:av}). The components
$\bar{r}_j$ of $\bar{\vek{r}}$ do not commute:
\begin{subequations}
  \begin{equation}
    \label{eq:dirac:pos:cent:com}
    [\bar{r}_j, \bar{r}_k] =
    -i \frac{\hbar c^2}{\Lambda_D^2} \, \eps_{jkl} \bar{S}_l \quad ,
  \end{equation}
  where
  \begin{equation}
    \label{eq:dirac:spin:cent}
    \bar{\vek{S}} = \vek{S}
      - \vek{F} \times \vek{p} \: \frac{\hbar}{\hat{\omega} (\vek{p})}
  \end{equation}
  is the spin operator analogous to Eq.\ (\ref{eq:gen:vel:av}) that
  does not mix the subspaces of positive and negative energy states.
  Diagonalizing Eq.\ (\ref{eq:dirac:pos:cent:com}) yields the
  uncertainty relation for $j \ne k$
  \begin{equation}
    \label{eq:dirac:pos:cent:unc}
    \Delta\bar{r}_j\, \Delta\bar{r}_k \ge
    \frac{\hbar^2 \, c^2}{4} \:
    \frac{\sqrt{m^2c^4 + |\eps_{jkl}| \, c^2p_l^2}}{\Lambda_D^3} \quad .
  \end{equation}
  Thus we obtain in the nonrelativistic limit (``$v \ll c$'')
  \begin{equation}
    \label{eq:dirac:pos:cent:unc0}
    \Delta\bar{r}_j\, \Delta\bar{r}_k \gtrsim
    \frack{1}{4} \, \lambdabar_D^2 \quad .
  \end{equation}
  In the opposite (ultrarelativistic) limit (``$v \lesssim c$'') we have
  \begin{equation}
    \label{eq:dirac:pos:cent:unc1}
    \Delta\bar{r}_j\, \Delta\bar{r}_k \gtrsim
    \frack{1}{4} \, \lambdabar_B^2 \quad .
  \end{equation}
\end{subequations}

The time derivative $\bar{\vek{v}}$ of the mean position operator
$\bar{\vek{r}}$ is the velocity operator one would expect based on
the correspondence principle and classical relativistic kinematics.
Its components $\bar{v}_j$, $\bar{v}_k$ commute, $[\bar{v}_j,
\bar{v}_k] = 0$. Furthermore, $\bar{\vek{v}}$ commutes with $H_D$,
i.e., it is a constant of the motion.

Equation (\ref{eq:dirac:pos:proj}) is motivated by the requirement
that it leaves the subspaces $\mathcal{H}_\pm$ separately invariant.
However, this requirement is not sufficient for a unique definition
of a relativistic position operator. The Dirac equation becomes
diagonal in the Foldy-Wouthuysen (FW) representation \cite{fol50}.
If we require the mean position operator to be diagonal in this
representation, it can be obtained in the standard representation
via an inverse FW transform:~\cite{tha92}
\begin{widetext}
  \begin{equation}
    \label{eq:dirac:pos:nw}
    \bar{\vek{r}}_{\SSs\mathrm{NW}} (t) =
    \vek{r} + \bar{\vek{v}} t
    - \frac{\hbar \beta}{2i\Lambda_D} \left[ c\vek{\alpha} -
      \frac{c^3 \left(\vek{\alpha} \cdot \vek{p}\right) \vek{p}}%
      {\Lambda_D(\Lambda_D+mc^2)} \right]
    - \frac{c^2 \, \vek{S} \times \vek{p}}{\Lambda_D(\Lambda_D+mc^2)}
    \quad .
  \end{equation}
\end{widetext}
This operator is often called the Newton-Wigner position operator
\cite{new49}. In contrast with the position operator $\bar{\vek{r}}$
in Eq.\ (\ref{eq:dirac:pos:proj}), the components of
$\bar{\vek{r}}_{\SSs\mathrm{NW}}$ \emph{do} commute. We see that
$\bar{\vek{r}}_{\SSs\mathrm{NW}} (t)$ shares with $\bar{\vek{r}}
(t)$ from Eq.\ (\ref{eq:dirac:pos:proj}) that the time derivative
$\bar{\vek{v}} (t)$ is given by Eq.\ (\ref{eq:gen:vel:av}),
i.e., it does not show any \emph{zitterbewegung\/} because it is a
constant of the motion.

General requirements for any position observable describing the
localization of a particle or wave packet are discussed in
Ref.~~\onlinecite{tha92}. In this context, the operator
$\bar{\vek{r}}$ in Eq.~(\ref{eq:dirac:pos:proj}) appears
inappropriate because its components do not commute. The optimal
choice for a position observable is the operator
$\bar{\vek{r}}_{\SSs\mathrm{NW}}$. However, general arguments
prohibit the possibility of strict spatial localization for a
one-particle state (see, e.g., Refs.~~\onlinecite{heg74, tha92,
deb06}). This imposes restrictions on the utility of any
position operator in relativistic systems.

\section{Rashba (and Pauli) Hamiltonian}
\label{sec:rash}

An intriguing example of \emph{zitterbewegung\/}-like dynamics
exhibited by a non-Dirac-like Hamiltonian has been
found~\cite{sch05a} in the Rashba model~\cite{byc84}. This model
describes 2D electrons in semiconductor heterostructures with
spin-orbit coupling present, using the effective Hamiltonian (we
assume here $B =0$, see Sec.~\ref{sec:lanrash} for the case $B\ne
0$)
\begin{subequations}
  \label{eq:rash:ham}
  \begin{equation}
    \label{eq:rash:ham0}
    H = \frac{p_x^2 + p_y^2}{2m} + H_R \quad .
  \end{equation}
  Here
  \begin{equation}
    \label{eq:rash:ham1}
    H_R = \alpha \, ( \vek{\sigma} \times \vek{p} ) \cdot \vek{e}_z
    = \alpha \left(\begin{array}{cc}
        0 & p_y + ip_x \\ p_y - ip_x & 0 \end{array}\right)
  \end{equation}
\end{subequations}
is the Rashba term with Rashba coefficient $\alpha$
(Ref.~~\onlinecite{byc84}), and $\vek{e}_z$ denotes the unit vector
in the direction perpendicular to the 2D plane. (Note that $H_R^2 =
\alpha^2 p^2$, similar to the Dirac Hamiltonian.) The Hamiltonian
(\ref{eq:rash:ham}) is also equivalent to the Pauli Hamiltonian
\cite{sak67} for a 2D system. The energy eigenvalues of $H$ are
\begin{equation}
  \label{eq:rash:en}
  E_\pm (\vek{p}) = \frac{p^2}{2m} \pm \alpha p \quad .
\end{equation}

The time-dependent position operator $\vek{r}(t)$ in the Rashba
model was discussed previously in Ref.~~\onlinecite{sch05a}.
Evaluated in close analogy to the Dirac case, it is again possible
to decompose $\vek{r}(t)$ into a mean part $\bar{\vek{r}} (t)$ and
an oscillating part $\tilde{\vek{r}} (t)$. The result is of the form
shown in Eq.~(\ref{eq:gen:posH}), where $\vek{F}$ and $\hat{\omega}
(\vek{p})$ are now
\begin{equation}
  \label{eq:rash:oF}
  \vek{F} = \frac{\partial H_R}{\partial \vek{p}}
  - \frac{\alpha^2 \vek{p}}{H_R}
  = \sigma_z \, \frac{\alpha^2 \, \vek{e}_z \times \vek{p}}{i H_R}
  \; , \hspace{2em}
  \hat{\omega} (\vek{p}) = \frac{2 H_R}{\hbar} \; .
\end{equation}
Explicit evaluation shows that $\vek{r}(t)$ oscillates with the
frequency $\omega_R = 2 \alpha p / \hbar$, which is equal to the
precession frequency of a spin moving in the effective magnetic
field of the Rashba term [see Eq.\ (\ref{eq:rash:spinH}) below]. The
oscillation becomes arbitrarily slow for $p \rightarrow 0$. We find
for the oscillating part of $\vek{r}(t)$
\begin{equation}
  \label{eq:rash:pos:osz}
  \tilde{r}^2 (t) = \lambdabar_B^2 / 4 \quad ,
\end{equation}
i.e., the magnitude of the oscillations is of the order of the
de~Broglie wave length $\lambdabar_B$ and independent of the Rashba
coefficient $\alpha$. Note that $\lambdabar_B$ diverges in the limit
$p \rightarrow 0$.

We obtain the mean part $\bar{\vek{r}} (t)$ by projecting on the
subspaces of $H$ associated with the spin-split bands, as in Eq.\
(\ref{eq:dirac:pos:proj}). We find the same $\bar{\vek{r}} (t)$ by
applying an inverse FW transformation, similar to Eq.\
(\ref{eq:dirac:pos:nw}). For the Rashba model, the last term in Eq.\
(\ref{eq:gen:pos:av}) corresponds to a spatial separation of up and
down spin contributions in a wave packet by $\sim \lambdabar_B$
(independent of the Rashba coefficient $\alpha$), which was noticed
in previous numerical work~\cite{bru05}. The general validity of
Eq.~(\ref{eq:gen:pos:av}) for two-band models implies the existence
of similar displacements for the Dirac and Luttinger cases. See also
Ref.~~\onlinecite{cse06}. The components $\bar{x}$ and $\bar{y}$ of
the mean position operator $\bar{\vek{r}}$ commute, similar to
$\bar{\vek{r}}_{\SSs\mathrm{NW}}$ in Eq.~(\ref{eq:dirac:pos:nw}).

The velocity operator and its derivative are given by
Eqs.~(\ref{eq:gen:vel}) and (\ref{eq:gen:accH}), respectively,
using expressions (\ref{eq:rash:oF}). The oscillatory part of
$\vek{v}$ satisfies
\begin{subequations}
  \label{eq:rash:vel:comunc}
  \begin{equation}
    \label{eq:rash:vel:mag}
    \tilde{v}^2 (t) = \alpha^2 \quad ,
  \end{equation}
  i.e., the magnitude of the oscillatory motion $\tilde{\vek{v}}
  (t)$ is given by the Rashba coefficient $\alpha$. On the other
  hand, the components of $\vek{v}$ do not commute, and we have
  \begin{equation}
    \label{eq:rash:vel:com}
      [v_x (t), v_y (t)] =
      2 i \alpha^2 \, \sigma_z \, \ee^{-i \hat{\omega} (\vek{p}) t} \quad ,
  \end{equation}
  which implies
  \begin{equation}
    \label{eq:rash:vel:unc}
    \Delta v_x \, \Delta v_y \ge \alpha^2
    = (\frack{1}{2} \omega_R \lambdabar_B)^2 \quad ,
  \end{equation}
\end{subequations}
analogous to Eqs.\ (\ref{eq:dirac:vel:comunc}). In Eq.\
(\ref{eq:rash:vel:unc}) we replaced the matrix-valued RHS of Eq.\
(\ref{eq:rash:vel:com}) by the eigenvalues of this matrix. Thus
similar to the Dirac case, both the magnitude of the oscillations in
$\vek{v}(t)$ and the minimum uncertainty are given by the same
parameter. The components of the mean part of the velocity operator
commute, $[\bar{v}_x (t), \bar{v}_y (t)] = 0$. They also commute
with $H_R$, i.e., they are constants of the motion.

The time dependence of orbital angular momentum $L_z$, spin
component $S_z$, and total angular momentum $J_z = L_z + S_z$ can be
straightforwardly discussed. We get
\begin{subequations}
\label{eq:rash:angH}
\begin{eqnarray}
   L_z (t) & = & L_z + \frac{\hbar\, \sigma_z}{2}
    \left(1 - \ee^{-i \hat{\omega} (\vek{p}) t} \right) \, ,
    \label{eq:rash:orbangH} \\
   S_z (t) & = & \frac{\hbar\, \sigma_z}{2} \:
    \ee^{-i \hat{\omega} (\vek{p}) t} \,\, , \label{eq:rash:spinH} \\
   J_z (t) & = & J_z = L_z + S_z \quad .
\end{eqnarray}
\end{subequations}
The formal structure of these equations is analogous to the
Dirac-case counterparts shown in Eqs.\ (\ref{eq:gen:angH}).
Equation~(\ref{eq:rash:spinH}) represents the well-known spin
precession in the effective magnetic field of the Rashba term, which
has been observed experimentally~\cite{cro05}. The total angular
momentum component perpendicular to the plane does not depend on
time, as expected from $[J_z, H] = 0$. Obviously Eqs.\
(\ref{eq:rash:angH}) require that the spin precession is caused by
the \emph{effective} in-plane magnetic field of a spin-orbit
coupling term such as the Rashba term. We see here clearly the
difference between spin precession caused by spin-orbit coupling and
spin precession caused by the Zeeman term in the presence of an
external in-plane magnetic field. In the latter case $J_z$ is not a
constant of motion.

\section{Luttinger Hamiltonian}
\label{sec:lutt}

The uppermost valence band $\Gamma_8^v$ of common semiconductors
like Ge and GaAs is well-characterized by the Luttinger Hamiltonian
\cite{lut56}
\begin{subequations}
  \begin{equation}
    \label{eq:lutt}
    H = - \frac{\gamma_1 \, p^2}{2m} + H_L \quad .
  \end{equation}
  We assume $B = 0$ and use the spherical approximation~\cite{lip70}
  \begin{equation}
    \label{eq:lutt:ham}
    H_L = \frac{\bar{\gamma}}{m} \left[
      \left(\vek{p} \cdot \vek{S} \right)^2 - \frack{5}{4} \,
      p^2 \openone_{4\times 4} \right] \quad ,
  \end{equation}
\end{subequations}
where $\gamma_1$ and $\bar{\gamma}$ are the dimensionless Luttinger
parameters, and $\vek{S}$ is the vector of $4 \times 4$ spin
matrices for a system with spin $s=3/2$. (Note $H_L^2 =
\bar{\gamma}^2 p^4 / m^2$, similar to the Dirac Hamiltonian.) The
twofold-degenerate energy eigenvalues of $H$ are
\begin{equation}
  \label{eq:lutt:en}
  E_\pm (\vek{p}) = - \frac{p^2}{2m} \:(\gamma_1 \pm 2 \bar{\gamma})
  \quad .
\end{equation}
The upper sign corresponds to the so-called light-hole (LH) states
with spin-$z$ component $M=\pm 1/2$, and the lower sign corresponds
to the heavy-hole (HH) states with $M=\pm 3/2$. The
momentum-dependent energy gap between HH and LH states is
\begin{equation}
  \hbar\omega_L = 2 \bar{\gamma} p^2 / m \quad .
\end{equation}
The position operator is of the form shown in
Eq.~(\ref{eq:gen:posH}) with $\vek{F}$ and $\hat{\omega} (\vek{p})$
given by
\begin{equation}
  \label{eq:lutt:oF}
  \vek{F} = \frac{\, \partial H_L}{\partial\vek{p}}
  - \frac{2\vek{p} \, H_L}{p^2} \; ,
  \hspace{2em}
  \hat{\omega} (\vek{p}) = \frac{2 H_L}{\hbar} \quad .
\end{equation}
Thus $\vek{r}(t)$ oscillates with the frequency $\omega_L$, which
has been noticed in previous numerical work~\cite{jia05}. Similar to
the Rashba Hamiltonian, these oscillations become arbitrarily slow
for $p \rightarrow 0$. The squared amplitude of the oscillations of
$\vek{r}(t)$ is $\tilde{r}^2(t) = (3/2) \lambdabar_B^2$, independent
of the Luttinger parameter $\bar{\gamma}$. It diverges for $p
\rightarrow 0$.

We obtain the mean position operator $\bar{\vek{r}}$, defined in
Eq.\ (\ref{eq:dirac:pos:proj}), using projection operators that
project onto HH and LH states \cite{mur04}. The result coincides
with Eq.\ (\ref{eq:gen:pos:av}). The components of $\bar{\vek{r}}$
do not commute,
\begin{equation}
  \label{eq:lutt:pos:com}
  [\bar{r}_j, \bar{r}_k] = \left[
    \frac{\hbar \, \partial H_L / \partial p_j}{2 \,H_L} ,
    \frac{\hbar \, \partial H_L / \partial p_k}{2 \,H_L} \right] \quad ,
\end{equation}
implying the uncertainty relation
\begin{equation}
  \label{eq:lutt:pos:unc}
  \Delta\bar{r}_j \, \Delta\bar{r}_k \ge
  \frac{3 \, \eps_{jkl} \, \hbar^2 p_l}{4 \, p^3} \quad .
\end{equation}
This uncertainty is of the order of (or less than) the de~Broglie
wave length. The uncertainty is the largest for those components
$\bar{r}_j$ that are perpendicular to $\vek{p}$.

Using Eqs.~(\ref{eq:lutt:oF}), the velocity operator can be written
in the form shown in Eq.~(\ref{eq:gen:vel}). For its oscillating
part, we find $\tilde{v}^2 (t) = 6 \bar{\gamma}^2 (p / m)^2$. The
components of $\vek{v}$ do not commute,
\begin{subequations}
  \begin{equation}
    \label{eq:lutt:vel:com}
    [v_j, v_k] = \left[ \frac{\partial H_L}{\partial p_j} ,
                        \frac{\partial H_L}{\partial p_k} \right] \quad ,
  \end{equation}
  which corresponds to the uncertainty relation
  \begin{equation}
    \label{eq:lutt:vel:unc}
    \Delta v_x\, \Delta v_y \ge \frac{\bar{\gamma}^2 \, p}{m^2} \:
    \sqrt{3 (4 p_x^2 + 4p_y^2 + 3p_z^2)}
  \end{equation}
\end{subequations}
and cyclic permutations thereof, i.e., the uncertainty is
approximately limited by $3\bar{\gamma}^2 \, (p/m)^2 = \frack{3}{4}
(\omega_L \, \lambdabar_B)^2$. Thus again, the magnitude of the
oscillations of $\vek{v}(t)$ and the minimum uncertainty are
characterized by the same combination of parameters. The velocity
$\vek{v} (t)$ is not a conserved quantity but satisfies
Eq.~(\ref{eq:gen:accH}). However, the mean velocity operator is
again given by Eq.\ (\ref{eq:gen:vel:av}). Its components commute
and are constants of the motion.

The time dependence of orbital angular momentum $\vek{L}$, spin
$\vek{S}$, and total angular momentum $\vek{J} = \vek{L} + \vek{S}$
turns out to be given by Eqs.\ (\ref{eq:gen:angH}). Note that the
time dependence of $\vek{S}(t)$ in the Luttinger model corresponds
to a spin precession in the absence of any external or effective
magnetic field \cite{cul06}. Again, the total angular momentum does
not depend on time, which reflects the fact that $[\vek{J}, H] = 0$.

We remark that a similar analysis as presented in this Section also
applies to models that neglect the spin degree of freedom. An
example is the $3\times 3$ Shockley Hamiltonian that describes
spinless holes in the uppermost valence band $\Gamma_5^v$ of
semiconductors like Si \cite{sho50,dre55}. Indeed, this is
consistent with the fact that \emph{zitterbewegung\/} for the Dirac
case can be studied already in a model with only one spatial
dimension, where the Dirac Hamiltonian $H_D$ becomes a $2\times 2$
matrix that reflects the occurence of both signs of the energy in
the spectrum of $H_D$; but this Hamiltonian does not describe the
spin degree of freedom \cite{tha04}. A spin with spin-orbit coupling
is not a necessary condition for the oscillatory behavior of
$\vek{r}(t)$ and $\vek{v}(t)$ to occur. The most basic ingredient
required for \emph{zitterbewegung\/}-like effects are several bands
separated by a (usually momentum-dependent) gap. Often the splitting
of these bands can be described by an \emph{effective} spin-orbit
coupling \cite{lip70,ber06}.

\section{Kane Hamiltonian}
\label{sec:kane}

\begin{figure}[tbp]
 \includegraphics[width=0.5\columnwidth]{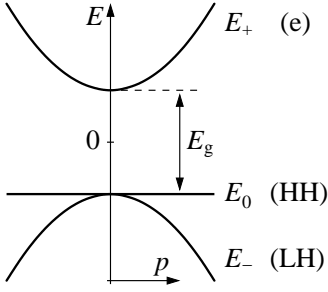}
 \caption{\label{fig:kane}Energy spectrum $E_{\pm,0} (\vek{p})$ of
 the $6\times 6$ Kane model. Here, $E_g$ denotes the fundamental
 gap. Each band $E (\vek{p})$ is twofold degenerate.}
\end{figure}

The Kane Hamiltonian \cite{kan57} is an effective Hamiltonian that
captures the important physics of electrons and holes in narrow-gap
semiconductors like InSb. We restrict ourselves to the $6\times 6$
Kane model which includes the lowest conduction band $\Gamma_6^c$
and the uppermost valence band $\Gamma_8^v$ (3D, $B=0$), neglecting
the split-off valence band $\Gamma_7^v$, because that model permits
a fully analytical solution. Then we have
\begin{equation}
  \label{eq:kane:ham}
  H_K = \left(\begin{array}{cs{1em}c}
        (E_g/2) \;\openone_{2\times 2}
      & \sqrt{3} \mathcal{P} \, \vek{T}\cdot\pp \\[1ex]
        \sqrt{3} \mathcal{P} \, \vek{T}^\dagger \cdot\pp
      & - (E_g/2) \; \openone_{4\times 4}
    \end{array}\right) \quad .
\end{equation}
Here $E_g$ is the fundamental energy gap, and $\mathcal{P}$ denotes
Kane's momentum matrix element. The vector $\vek{T}$ of $2\times 4$
matrices is defined in Ref.~~\onlinecite{win03}. The energy
eigenvalues of $H_K$ are (each twofold degenerate)
\begin{subequations}
  \begin{equation}
    \label{eq:kane:en}
    E_\pm (\pp) = \pm \Lambda_K \, , \qquad
    E_0 (\pp) =  - E_g/2 \, ,
  \end{equation}
  where
  \begin{equation}
    \label{eq:kane:lambda}
    \Lambda_K = \sqrt{(E_g/2)^2 + \frack{2}{3}\mathcal{P}^2 p^2}  \quad ,
  \end{equation}
\end{subequations}
i.e., the Kane Hamiltonian combines the gapped spectrum of the Dirac
Hamiltonian with the gapless spectrum of the Luttinger Hamiltonian.
(Indeed, the Luttinger Hamiltonian corresponds to the limiting case
$E_g \rightarrow \infty$ of the Kane model.) The energy spectrum
$E_{\pm,0} (p)$ is shown in Fig.~\ref{fig:kane}.

Results similar to those discussed below can also be derived
perturbatively for the full $8\times 8$ Kane Hamiltonian that
includes the split-off valence band $\Gamma_7^v$. We also remark
that a simplified $4 \times 4$ Kane Hamiltonian, which includes only
the conduction band $\Gamma_6^c$ and the valence band $\Gamma_7^v$,
is strictly equivalent to the Dirac Hamiltonian. Recently,
\emph{zitterbewegung\/} was studied for a simplified version of the
Kane model where the HH band [with dispersion $E_0 (p) = -E_g/2$]
and the split-off band $\Gamma_7^v$ were neglected \cite{zaw05a}. In
this limit, the Kane Hamiltonian becomes similar to the Dirac
Hamiltonian. Our analysis below shows that qualitatively new aspects
arise when the HH band is taken into account.

Similar to the Dirac equation, the velocity operator $\vek{v}$ in
the Kane model has a discrete spectrum. Each component of $\vek{v}$
has eigenvalues $\pm \sqrt{2/3} \, \mathcal{P}$ and $0$, which
correspond to (pure) electron, LH, and HH states. In general, for a
wave packet containing a superposition of electron, LH, and HH
states we have a finite probability to measure each of these
discrete values.
The components of the velocity $\vek{v}$ do not commute. We get
\begin{subequations}
  \begin{equation}
    \label{eq:kane:cvel}
    [v_j, v_k] = - \frac{2i}{3} \, \mathcal{P}^2
    \, \eps_{jkl} \left(\begin{array}{cc}
        \sigma_l & 0 \\ 0 & \Sigma_l \end{array} \right) \quad ,
  \end{equation}
  where $\Sigma_l$ are the $4\times 4$ spin matrices for spin
  $s=3/2$. This corresponds to the uncertainty relation for $j\ne k$
  \begin{equation}
    \label{eq:kanevel:unc}
    \Delta v_j \, \Delta v_k \ge \frac{\mathcal{P}^2}{6} \quad .
  \end{equation}
\end{subequations}
Note that Eq.\ (\ref{eq:kane:cvel}) implies that the minimum
uncertainty depends on the dominant character of the wave function.
The lower bound $\mathcal{P}^2 / 6$ requires a LH state. For an
electron state, the minimum uncertainty is $\mathcal{P}^2 / 3$,
whereas for a HH state it is $\mathcal{P}^2 / 2$.

We omit here the lengthy expressions for $\vek{r}(t)$ and
$\vek{v}(t)$. It follows from Eq.\ (\ref{eq:kane:en}) that the
oscillating parts $\tilde{\vek{r}} (t)$ and $\tilde{\vek{v}} (t)$ of
$\vek{r}(t)$ and $\vek{v}(t)$ depend on the frequencies
\begin{subequations}
  \label{eq:kane:freq}
  \begin{eqnarray}
    \hbar\omega_{+-} & \equiv & E_+ - E_-
    = 2 \Lambda_K \quad , \\[1ex]
    \hbar\omega_{\pm0} & \equiv & E_\pm - E_0
    = E_g/2 \pm \Lambda_K \quad .
  \end{eqnarray}
\end{subequations}
Unlike in the models discussed above, $\tilde{r}^2 (t)$ and
$\tilde{v}^2 (t)$ in the Kane model are not diagonal in spin space.
Hence these quantities depend explicitly on time, oscillating with
the frequencies given in Eqs.\ (\ref{eq:kane:freq}).
However, we can estimate the magnitude of these quantities by
neglecting the oscillatory terms and diagonalizing the resulting
matrices. We get the following twofold-degenerate eigenvalues
for~$\tilde{r}^2 (t)$
\begin{equation}
  \label{eq:kane:poso}
  \tilde{r}^2 (t) \simeq \left\{ \begin{array}{l}
      \displaystyle \frac{3\, \hbar^2}{2 \, p^2} \quad , \\[4ex]
      \displaystyle \left(\frac{7 \Lambda_K^4 + E_g^2 \Lambda_K^2 / 4
        - E_g^4/8}{8 \Lambda_K^4}
        \pm \frac{3\, E_g}{8 \Lambda_K} \right)
      \frac{\hbar^2}{p^2} \, .
  \end{array} \right.
\end{equation}
A Taylor expansion shows that for small mean velocities
(``nonrelativistic limit'') we thus have two characteristic length
scales for the oscillatory motion, the de Broglie wave length
$\lambdabar_B$ and an effective Compton wave length \cite{zaw05a}
\begin{equation}
  \label{eq:kane:compton}
  \lambdabar_K \equiv \frac{\hbar \mathcal{P}}{E_g} \quad .
\end{equation}
We have $\lambdabar_K \simeq 7$~{\AA} in GaAs and $\lambdabar_K
\simeq 40$~{\AA} in InSb which should be compared with $\lambdabar_D
= 3.9 \times 10^{-3}$~{\AA}. Note that, in the nonrelativistic
limit, the de Broglie wave length becomes a \emph{fourfold}
degenerate eigenvalue of $\tilde{r}^2 (t)$, i.e., it characterizes
the oscillatory motion of electron, HH, and LH states.
For large mean velocities (``relativistic limit''), the de Broglie
wave length is the only length scale characterizing $\tilde{r} (t)$.
Similarly, we get for $\tilde{v}^2 (t)$
\begin{equation}
  \label{eq:kane:velo}
  \tilde{v}^2 (t) \simeq \left\{ \begin{array}{l}
      \mathcal{P}^2 \\[1ex]
      \displaystyle \mathcal{P}^2  \left(\frack{5}{6} \Lambda_K^2
        + \frack{1}{6} E_g^2 \pm \frack{1}{4} E_g \Lambda_K \right)
      /\Lambda_K^2 \quad ,
    \end{array} \right.
\end{equation}
i.e., the magnitude of $\tilde{v}$ is of the order of $\mathcal{P}$
for both small and large mean velocities. Again, the minimum
uncertainty of $\vek{v}$ [Eq.\ (\ref{eq:kanevel:unc})] and the
magnitude of the oscillations of $\vek{v}$ are characterized by the
same parameter.

The mean velocity reads
\begin{equation}
  \label{eq:kane:velHm}
  \bar{\vek{v}} (t) = \left( H_K + \frac{E_g}{2} \right)
  \left(1 - \frac{H_K E_g}{2 \Lambda_K^2} \right)
  \frac{\vek{p}}{p^2} \; .
\end{equation}
The components of $\bar{\vek{v}}$ commute with each other and they
are constants of the motion.
The mean position operator reads
\begin{subequations}
\begin{widetext}
\begin{equation}
  \label{eq:kane:posHm}
  \bar{\vek{r}} (t) = \vek{r} + \bar{\vek{v}} t
  + \left\{ 1 - \frac{3E_g \, H_K}{p^2 \mathcal{P}^2} ,
           \frac{\hbar^2 \, \dot{\vek{v}}}{4 \Lambda_K^2} \right\}
  + \left[ \frac{\partial \tilde{H}_K^2}{\partial \vek{p}}
         - \frac{2 \vek{p} \tilde{H}_K^2}{p^2} \right]
    \frac{\hbar \, [3 \Lambda_K^2 + (E_g/2)^2]}
    {8i \, \tilde{H}_K^2 \Lambda_K^2} \; ,
\end{equation}
\end{widetext}
where $\{ A, B \} = \frack{1}{2} (A B + B A )$ denotes the
symmetrized product of $A$ and $B$,
\begin{equation}
  \label{eq:kane:hamt}
  \tilde{H}_K^2 \equiv \bigg( \frac{E_g^2}{H_K^2} - 2 \bigg) \:
       \bigg( \Lambda_K^2 - \frac{E_g^2}{2} \bigg) \quad ,
\end{equation}
\end{subequations}
and $\dot{\vek{v}}$ denotes the acceleration $\dot{\vek{v}} =
(i/\hbar) [ H_K, \vek{v}]$ . The components of $\bar{\vek{r}} (t)$
do not commute with each other. We do not give here the lengthy
expressions.

Orbital angular momentum $\vek{L} = \vek{r} \times \vek{p}$ and spin
$\vek{S}$ also oscillate as a function of time. Similar to the Dirac
and Luttinger cases, these oscillations arise even though free
particles are considered with no external or effective magnetic
field present. However, the total angular momentum $\vek{J} =
\vek{L} + \vek{S}$ does not oscillate as a function of time which,
as always, reflects the fact that $[\vek{J}, H_K] =0$.

\section{Landau Hamiltonian}
\label{sec:lan}

There are several remarkable similarities between the spin-dependent
Hamiltonians discussed above and the well-known and rather simple
case of the Landau Hamiltonian~\cite{lan30} describing the cyclotron
motion of 2D electrons in the presence of a magnetic field $B_z > 0$
perpendicular to the 2D plane. The Landau Hamiltonian is given by
\begin{equation}
  \label{eq:lan}
  H_c = \frac{p_x^2 + p_y^2}{2m} \quad ,
\end{equation}
where $\vek{p}$ is the kinetic momentum with
\begin{equation}
  \label{eq:lan:pos:com}
  [p_x, p_y] = - i\, \hbar\, e B_z \quad .
\end{equation}
For the elementary charge $e$ we use the convention $e = |e|$. The
time-dependent position operator
\begin{subequations}
\begin{equation}
  \label{eq:lan:post}
  \vek{r} (t) = \vek{r} + \frac{\vek{p}}{m \omega_c}  \sin (\omega_c t)
  + \frac{\vek{p} \times \vek{e}_z}{m \omega_c} \left[ \cos (\omega_c t)
   - 1\right]
\end{equation}
can be written in a compact form using the complex notation $R = x -
i y$ and $P = p_x - i p_y$ (Ref.~~\onlinecite{mac95}), which
highlights the analogies between the Landau Hamiltonian and the
models in the preceding sections. We get:
\begin{equation}
  \label{eq:complexPos}
  R(t) = R + \frac{P}{m} \, \frac{1 - \ee^{- i \omega_c t}}{i \omega_c} \quad ,
\end{equation}
\end{subequations}
where $\omega_c = e B_z / m$ is the cyclotron frequency. Equation
(\ref{eq:complexPos}) shows that $P/m$ behaves similar to the
$\vek{F}$ operators in the preceding sections~\cite{Pm}. The
magnitude of the oscillations of $R(t)$ is the radius $\Lambda_c =
p/(m \omega_c)$ of the cyclotron orbit. Ignoring the oscillations
with frequency $\omega_c$, we have
\begin{equation}
  \label{eq:lan:pos:cent}
  \bar{R} (t) = R - \frac{i P}{m \omega_c} \equiv C
\end{equation}
independent of $t$, which corresponds to the center of the cyclotron
orbit (the guiding center). The components $\bar{x}$ and $\bar{y}$
of $\bar{R}$ do not commute
\begin{subequations}
  \label{eq:lan:pos:cent:cu}
  \begin{equation}
    \label{eq:lan:pos:cent:com}
    [\bar{x}, \bar{y}] = i \lambda_c^2 \quad ,
  \end{equation}
  where $\lambda_c = \sqrt{\hbar/ (e B_z)}$ is the magnetic length.
  Equation~(\ref{eq:lan:pos:cent:com}) can be written as an
  uncertainty relation
  \begin{equation}
    \label{eq:lan:pos:cent:unc}
    \Delta\bar{x} \, \Delta\bar{y} \ge \frack{1}{2} \, \lambda_c^2 \quad .
  \end{equation}
\end{subequations}
The velocity operator, in complex notation $V = v_x - i v_y$, is
given by
\begin{equation}
  \label{eq:lan:velt}
  V (t) = \frac{P}{m} \, \ee^{- i \omega_c t} \quad ,
\end{equation}
so that $\tilde{\vek{v}} (t) = \vek{v} (t)$ and $\tilde{v}^2 (t) =
(\omega_c \Lambda_c)^2$. The components $v_x$ and $v_y$ do not
commute,
\begin{subequations}
  \begin{equation}
    \label{eq:lan:vel:com}
    [v_x, v_y] = [v_x (t), v_y (t)] = - \frac{i\, \hbar\, e B_z}{m^2} \quad ,
  \end{equation}
  which corresponds to the uncertainty relation
  \begin{equation}
    \label{eq:lan:vel:unc}
    \Delta v_x \, \Delta v_y
    \ge \frack{1}{2} \: (\omega_c \, \lambda_c)^2 \quad ,
  \end{equation}
\end{subequations}
which should be compared with Eqs.\ (\ref{eq:dirac:vel:comunc}) and
(\ref{eq:rash:vel:comunc}). Obviously, implications arising from
this uncertainty relation become relevant only for sufficiently
large magnetic fields when $\lambda_c$ becomes comparable to
$\Lambda_c$.

The velocity $V(t)$ is not a conserved quantity, which reflects the
effect of the Lorentz force. We have
\begin{equation}
  \label{eq:lan:acc}
 \dot V (t) = \frac{- i \omega_c P}{m} \, \ee^{- i \omega_c t} \quad .
\end{equation}
The mean velocity operator vanishes,
\begin{equation}
  \label{eq:lan:vel:mean}
  \bar V = 0 \quad ,
\end{equation}
because, on average, the particle is at rest for $B_z \ne 0$.
This also implies $[\bar{v}_x, \bar{v}_y]  = 0$.

Our analysis indicates that the dynamical properties of
the Landau model bear strong resemblances to those exhibited by
models showing \emph{zitterbewegung\/}-like motion.

\section{Landau-Rashba Hamiltonian}
\label{sec:lanrash}

An interesting example combining two types of oscillatory motion can
be found by considering the interplay between 2D cyclotron motion
(Sec.~\ref{sec:lan}) and Rashba spin splitting
(Sec.~\ref{sec:rash}). The Hamiltonian for that situation reads
\begin{equation}
  H_{cR} = H_c + H_R + \frac{g}{2}\, \mu_{\text{B}} \sigma_z B_z \quad .
\end{equation}
Here we have also included a Zeeman term with Land\'e factor $g$ and
Bohr magnetic moment $\mu_{\text{B}} = e\hbar/ (2m_e)$ (where
$m_{\text{e}}$ denotes the electron mass in vacuum), and the terms
$H_R$ and $H_c$ are given in Eqs.\ (\ref{eq:rash:ham1}) and
(\ref{eq:lan}). For the following calculation we replace the
components $p_x$ and $p_y$ of the kinetic momentum by creation and
annihilation operation operators for Landau levels, $a^\dagger$ and
$a$, defined in the usual way,
\begin{equation} \label{eq:aDef}
  a = \frac{\lambda_c \, P}{\sqrt{2}\hbar} \quad ,
\end{equation}
and $a^\dagger$ is the adjoint of $a$. The resulting expression for
$H_{cR}$ (Ref.~~\onlinecite{ras60}) is equivalent to the
Jaynes-Cummings model~\cite{sho93} in the rotating-wave
approximation. To find the time evolution of the observables in the
Heisenberg picture, we first separate $H_{cR}$ into two commuting
parts, $H_{cR} = H_{cR}^{(1)} + H_{cR}^{(2)}$, where
\begin{subequations}
  \begin{eqnarray}
    H_{cR}^{(1)} &=& \hbar \omega_c \left( a^\dagger a + \frac{1 +
     \sigma_z}{2} \right) \quad , \\
    H_{cR}^{(2)} &=& \frac{\sqrt{2} i \hbar \, \alpha}{\lambda_c}
    \left ( a \sigma_+ - a^\dagger \sigma_- \right)
    - \frac{\hbar \omega_c}{2} \,
    \left( 1 - \frac{g m}{2 m_{\text{e}}} \right)\sigma_z \quad .
    \nonumber \\
  \end{eqnarray}
\end{subequations}
Here we used $\sigma_\pm \equiv (\sigma_x \pm i \sigma_y)/2$.

It is straightforward to calculate the time evolution of the spin
component parallel to the magnetic field,
\begin{equation}
  \label{eq:rashlan:vel:szPrecess}
  S_z(t) = S_z - i \sigma_z  H_R \,
  		\frac{1 - \ee^{-2 i H_{cR}^{(2)} t /\hbar}}
		{2 i H_{cR}^{(2)} /\hbar} \quad .
\end{equation}
This result is the generalization of Eq.\ (\ref{eq:rash:spinH}) to
the case of a finite magnetic field. Interestingly, time averaging
the r.h.s of Eq.~(\ref{eq:rashlan:vel:szPrecess}) does not result in
a vanishing spin component parallel to the field direction. We find
\begin{subequations}
  \begin{equation}
    \bar{S}_z =  - \frac{\hbar^2 \omega_c}{4 \, H_{cR}^{(2)}}
    			\left( 1 - \frac{g m}{2 m_{\text{e}}} \right) \quad .
  \end{equation}
  Neglecting Zeeman splitting and considering the limit of small
  $B_z$, this result becomes
  \begin{equation}
    \bar{S}_z \approx - \frac{\hbar^2 \omega_c}{4 \alpha p^2}\,
    (\vek{\sigma} \times \vek{p} ) \cdot \vek{e}_z \quad ,
  \end{equation}
\end{subequations}
which is exactly the finite value of the spin component parallel to
the magnetic field that was obtained in semiclassical calculations
of spin-split cyclotron orbits. \cite{rey04}

To calculate the time evolution of the position operator, we use the
complex notation from Sec.~\ref{sec:lan}. We have
\begin{equation}
  R = C + \frac{i P}{m \omega_c} \equiv C + i \sqrt{2} \lambda_c \, a
   \quad ,
\end{equation}
where $C$ is the position of the guiding center, see Eq.\
(\ref{eq:lan:pos:cent}). Even in the presence of $H_R$, the guiding
center $C$ remains a constant of the motion, $[C, H_{cR}] = 0$. The
time evolution of $P$ due to $H_{cR}^{(1)}$ is just a trivial factor
$\ee^{- i \omega_c t}$, so that we only need to evaluate the time
evolution of $P \propto a$ under $H_{cR}^{(2)}$. (Note that
$[H_{cR}^{(1)}, H_{cR}^{(2)}] = 0$.) This problem has been solved
for the Jaynes-Cummings model~\cite{sho93,bar97}. Translating into
our situation, we get for the time-dependent position operator
\begin{subequations}
\begin{widetext}
  \begin{equation}
    \label{eq:rashlan:pos}
    R(t) = C
    + \frac{i \exp\, [ - i (\omega_c + \omega_+) t]} {\omega_- - \omega_+}
    \left(\frac{\omega_-}{\omega_c}\, \frac{P}{m} + 2 i \alpha \sigma_- \right)
    - \frac{i \exp\, [- i (\omega_c + \omega_-) t]}{\omega_- - \omega_+}
    \left(\frac{\omega_+}{\omega_c}\, \frac{P}{m} + 2 i \alpha \sigma_- \right)
    \quad ,
  \end{equation}
\end{widetext}
with the frequency operators $\omega_\pm$ given by
\begin{equation}
  \hbar\omega_\pm =  - H_{cR}^{(2)} \pm \sqrt{\big( H_{cR}^{(2)} \big)^2
  + 2 \hbar \omega_c m \alpha^2} \quad .
\end{equation}
\end{subequations}
The terms proportional to $\sigma_-$ in Eq.\ (\ref{eq:rashlan:pos})
are reminiscent of the oscillatory motion in the Rashba case for
$B_z=0$, where the amplitude of the oscillations is inversely
proportional to the de Broglie wave length and independent of
$\alpha$, see Eq.\ (\ref{eq:rash:pos:osz}). Here these terms
contribute to a spin-dependent renormalization of the cyclotron
radius. We also note that $\bar{R}(t) = C$, so that Eq.\
(\ref{eq:lan:pos:cent:cu}) remains valid in the presence of $H_R$.

The velocity operator is given by
\begin{equation}
  \label{eq:rashlan:vel}
  V \equiv \dot R = \frac{P}{m} - 2 i \alpha \sigma_-  \quad .
\end{equation}
The commutator of the components of $V$,
\begin{subequations}
  \begin{equation}
    \label{eq:rashlan:vel:com}
    [v_x, v_y]  = - \frac{i \hbar \, e B_z}{m^2} + 2 i \alpha^2 \sigma_z \quad ,
  \end{equation}
  is the sum of the corresponding results obtained separately from
  $H_c$ and $H_R$ [see Eqs.\ (\ref{eq:rash:vel:com}) and
  (\ref{eq:lan:vel:com})]. However, in the uncertainty relation
  \begin{equation}
    \label{eq:rashlan:vel:unc}
    \Delta v_x\, \Delta v_y \ge
    \left|\frac{\hbar \, e B_z}{2 m^2} - \alpha^2 \right| ,
  \end{equation}
\end{subequations}
the two contributions are subtracted, thus \emph{reducing\/} the
minimum uncertainty.
The time dependence of $V$ can be readily obtained by taking the
time derivative of Eq.\ (\ref{eq:rashlan:pos}). It can be written as
\begin{subequations}
  \begin{equation}
    V(t) = \ee^{-i (\omega_c + \omega_+) t} F_+
           + \ee^{-i (\omega_c + \omega_-) t} F_- \quad ,
  \end{equation}
  with the complex amplitude operators
  \begin{equation}
    F_\pm = \frac{V}{2} \pm \bigg( \frac{H_{cR}^{(2)} + 2 m \alpha^2}
            {\omega_+ - \omega_-}\frac{P}{m}
            + 2 i \alpha \, \frac{H_{cR}^{(2)} - \hbar \omega_c}
            {\omega_+ - \omega_-} \, \sigma_- \bigg) \, .
  \end{equation}
\end{subequations}

\section{Conclusions and Outlook}
\label{sec:concl}

We studied a variety of qualitatively different model Hamiltonians
for quasi-free electrons that exhibit \emph{zitterbewegung\/}-like
oscillatory motion. A number of features can be identified that are
widely shared as discussed in Sec.~\ref{sec:intro}. Here we finally
point out open questions.

For the Dirac Hamiltonian, the amplitude of the
\emph{zitterbewegung\/} of $\vek{r}(t)$ is given by the Compton
wavelength in the nonrelativistic limit and by the de Broglie wave
length in the relativistic limit. For those Hamiltonians having a
gap that vanishes for $p \rightarrow 0$, the length scale of
oscillations in $\vek{r}(t)$ is always given by the de Broglie
wavelength $\lambdabar_B = \hbar/p$, independent of the magnitude of
spin-orbit coupling. It is surprising that the amplitude of the
oscillations of $\vek{r}(t)$ diverges in the nonrelativistic limit
$p \rightarrow 0$.

The most interesting but also, at least in our present work, a
largely open aspect is the experimental observability of
\emph{zitterbewegung\/}-like effects. Certainly, any measurement of
the oscillatory motion must obey the fundamental uncertainty
relations [Eq.~(\ref{eq:gen:vel:unc})] discussed in our work.
Furthermore, we have already commented on the intimate relation
between oscillations in position and spin space. However, while spin
precession due to spin-orbit coupling can be observed experimentally
\cite{cro05}, it is often argued that the \emph{zitterbewegung\/} of
$\vek{r}(t)$ ``is not an observable motion, for any attempt to
determine the position of the electron to better than a Compton
wavelength must defeat its purpose by the creation of
electron-positron pairs'' (Ref.~~\onlinecite{hua52}). We note that
the same argument can be applied to Bloch electrons in solids where
electron-hole pairs can be created \cite{lur70}.

\begin{acknowledgments}

RW thanks D.~Culcer and UZ thanks P.~Schwerdtfeger and G.~Vignale
for enlightening discussions. UZ is supported by the Marsden Fund Council
from Government funding, administered by the Royal Society of New
Zealand. Research at KITP was supported by the National Science
Foundation under Grant No.\ PHY99-07949.

\end{acknowledgments}

\appendix*
\section{Important Formulae}

Here we briefly summarize important formulae that are used in our
discussion of the oscillatory motion in various models. The
Heisenberg equation of motion for an operator $A$ reads
\begin{equation}
  \label{eq:hom}
  \frac{dA}{dt} = \frac{i}{\hbar} [H, A] \quad .
\end{equation}
It has the formal solution
\begin{equation}
  \label{eq:hop}
  A(t) = \ee^{iHt/\hbar} \,A(0)\, \ee^{-iHt/\hbar} \quad .
\end{equation}
In particular, the velocity operator $\vek{v}$ is defined by the
Heisenberg equation of motion for the position operator~$\vek{r}$,
\begin{equation}
  \label{eq:vel}
  \vek{v} \equiv \frac{d\vek{r}}{dt} = \frac{i}{\hbar} [H, \vek{r}] \quad .
\end{equation}
Throughout we use the convention that $A(t)$ denotes an operator in
the Heisenberg picture and $A = A(0)$ is the corresponding operator
in the Schr\"odinger picture.

In general, the uncertainty principle for two noncommuting
observables $A$ and $B$ reads
\begin{equation}
  \label{eq:uncert:rel}
  \Delta A \: \Delta B \ge
  \frack{1}{2} \: |\expectk{[ A, B]}| \quad ,
\end{equation}
where the uncertainty $\Delta A$ of $A$ is defined as
\begin{equation}
  \label{eq:uncert}
  \Delta A \equiv \sqrt{\expectk{A^2} - \expectk{A}^2} \quad .
\end{equation}



\begin{thebibliography}{10}
\bibitem{2by2}
Similarly, the discussion in Refs.~~\onlinecite{fer90, zaw06, zaw05a,
zaw05b, kat06, shm95, lur70, sch05a} was restricted to effective
$2\times 2$ Hamiltonians.
\bibitem{Pm}
This is only a partial analogy, because $P/m$ acts as a
unidirectional ladder operator on the infinite set of Landau levels
[see Eq.\ (\ref{eq:aDef})], while the $\vek{F}$ operators map
states between two bands. However, note also that an alternative
sign convention for the complex notation is $P' = p_x + ip_y$ so
that $P' \propto a^\dagger$.
\end{thebibliography}
\end{document}